# Substrate effect on thermal conductivity of monolayer WS$_2$: Experimental measurement and theoretical analysis

*Yufeng Zhang[1, †], Qian Lv[2, †], Aoran Fan[1], Lingxiao Yu[2], Haidong Wang[1], Weigang Ma[1], Xing Zhang[1, *], Ruitao Lv[2, 3, *]*

[1] Key Laboratory for Thermal Science and Power Engineering of Ministry of Education, Department of Engineering Mechanics, Tsinghua University, Beijing 100084, China

[2] State Key Laboratory of New Ceramics and Fine Processing, School of Materials Science and Engineering, Tsinghua University, Beijing 100084, China

[3] Key Laboratory of Advanced Materials (MOE), School of Materials Science and Engineering, Tsinghua University, Beijing 100084, China.

ABSTRACT

Monolayer $WS_2$ has been a competitive candidate in electrical and optoelectronic devices due to its superior optoelectronic properties. To tackle the challenge of thermal management caused by the decreased size and concentrated heat in modern ICs, it is of great significance to accurately characterize the thermal conductivity of the monolayer $WS_2$, especially with substrate supported. In this work, the dual-wavelength flash Raman method is used to experimentally measure the thermal conductivity of the suspended and the $Si/SiO_2$ substrate supported monolayer $WS_2$ at a temperature range of 200 K - 400 K. The room-temperature thermal conductivity of suspended and supported $WS_2$ are 28.45±6.52 W/m·K and 15.39±4.96 W/m·K, respectively, with a ~50% reduction due to substrate effect. To systematically study the underlying mechanism behind the striking reduction, we employed the Raman spatial mapping analysis combined with the molecular dynamics simulation. The analysis of Raman spectra showed the increase of doping level, reduction of phonon lifetime and suppression of out-of-plane vibration mode due to substrate effect. In addition, the phonon transmission coefficient was mutually verified with Raman spectra analysis and further revealed that the substrate effect significantly enhances the phonon scattering at the interface and mainly suppresses the acoustic phonon, thus leading to the reduction of thermal conductivity. The thermal conductivity of other suspended and supported monolayer TMDCs (e.g. $MoS_2$, $MoSe_2$ and $WSe_2$) were also listed for comparison. Our researches can be extended to understand the substrate effect of other 2D TMDCs and provide guidance for future TMDCs-based electrical and optoelectronic devices.

INTRODUCTION

The researches of two-dimensional (2D) materials have been ignited since the discovery of graphene, a single-atom-thick carbon sheet [1]. Among many 2D materials, transition metal dichalcogenides (TMDCs) are widely recognized to be the most promising ones due to their fascinating properties [2, 3]. Unlike graphene with the zero bandgap, TMDCs exhibit the transition from an indirect band gap to a direct band gap with the decreasing layers, opening up various opportunities for the applications in electrical and optoelectronic devices [4-8]. Historically, the popularity of molybdenum dichalcogenides ($MoS_2$) has overshadowed the potentials of tungsten dichalcogenides ($WS_2$) to some extent. However, $WS_2$, especially the monolayer structure, has recently attracted considerable attention due to its high carrier mobility [9], large band gap [10], strong photoluminescence (PL) [8] and giant exciton binding energy [11]. Owing to these superior properties, $WS_2$ has been utilized in photodetectors [12], light-emitting diodes (LEDs) [13], optical modulators [14], memristors [15] and field effect transistors (FETs) [16], all of which exhibit excellent device performance. For example, monolayer $WS_2$-based photodetectors can realize a high photoresponsivity of 22.1 A $W^{-1}$ [17]. In addition, FETs based on monolayer $WS_2$ sandwiched between two BN layers show a large ON/OFF current ratio of $10^7$ and the carrier mobility can reach to 214 $cm^2$ $V^{-1}$ $s^{-1}$ in ambient [18].

As we know, FETs are the key components of modern integrated circuits (ICs). With the rapid development of design and processing technology, the characteristic length of transistors continuous to decrease, which brings new opportunities for 2D semiconductors due to their moderate band gap and ultrathin thickness [19, 20]. Among many candidates, monolayer $WS_2$ is predicted to be promising for high-performance transistors owing to its high electron mobility and large ON/OFF ratio [9, 18, 21-23]. However, the miniaturization of circuits can also cause many problems. The formation of local hot spots due to the substantial increase of power consumption per unit area may damage the circuits, which poses a great challenge to the thermal management of modern ICs. Based on this, the thermal properties of materials are urgent to be considered, which is equally important with electrical properties in IC design. Up to date, the experimental study on thermal properties of monolayer $WS_2$ is still lacking. The only experimental data of thermal conductivity of monolayer $WS_2$ are 32 W/m·K measured by Peimyoo et al. [24] and 20 (16) W/m·K measured by Vieira et al. [25] both using steady-state Raman method at room temperature. The large discrepancy calls for a benchmark thermal investigation of monolayer $WS_2$.

Steady-state Raman spectroscopy method has been widespread used in thermal conductivity determination of 2D materials due to its *in-situ* and non-invasive advantage [24-29]. However, several uncertainties originate from the fitting of temperature/power-dependent Raman peak shift and the determination of laser absorption coefficient could lead to significant measurement errors. It is important to notice that laser absorption coefficient is strongly dependent on the wavelength of incident laser, stress, defects, surface contamination and multiple reflections caused by substrate, so large variations are existed among samples at the nanoscale [30]. Take graphene as an example, different laser absorption coefficient can lead to quite scattered measurement data with large uncertainties [31, 32]. Therefore, the thermal conductivity of monolayer $WS_2$ still needs to be accurately measured.

Actually, modern 2D devices are usually fabricated on $Si/SiO_2$ insulating substrates to avoid electric leakage. Previous studies of graphene have reported that substrate can sharply influence the intrinsic thermal conductivity of single-layer graphene due to strong interface scattering [26, 33]. The similar phenomenon has been found in 2D TMDCs such as $MoS_2$, $MoSe_2$ and $WSe_2$, but the underlying mechanism has not been thoroughly studied [27, 29]. In addition, to meet the requirement of variable working conditions, the thermal conductivity versus temperature should be accurately measured. However, no experimental data on the thermal conductivity of monolayer TMDCs at different temperatures have been reported yet. Therefore, it is urgent to systematically study the influence of substrate and temperature on the thermal conductivity of TMDCs, which is of great significance to the thermal management of TMDCs-based devices.

In this work, we experimentally measured the thermal conductivity of suspended and supported monolayer $WS_2$ at 200 K-400 K by dual-wavelength flash Raman method [34, 35]. In addition, we employed molecular dynamics (MD) simulation to understand the mechanism of the substrate and temperature effect. The comparison of Raman spatial mapping between suspended and supported WS$_2$ reveal the influence of substrate from another aspect, and is mutually verified with the results of MD simulation.

RESULTS AND DISCUSSIONS

The monolayer WS$_2$ crystals were synthesized on SiO$_2$/Si substrate by an atmospheric-pressure chemical vapor deposition (AP-CVD) route. The optical image of WS$_2$ shows a triangular shape with the uniform contrast, demonstrating the uniform thickness of the crystal (Figure 1b). The crystalline structures of WS$_2$ crystals were investigated by transmission electron microscopy (TEM). The high-resolution TEM (HRTEM) image of WS$_2$ displays the six-fold symmetry honeycomb structure with the interplanar spacing distance of ~0.29 nm, assigned to the (110) plane of WS$_2$ [36]. The selected area electron diffraction (SAED) pattern only shows a set of hexagonally symmetric diffraction spots in Figure 1d, indicating the single crystal structure of WS$_2$. Raman spectrum and PL measurement were used to evaluate the crystalline quality and optical structures of WS$_2$ excited by a 532 nm laser, as represented in Figure 1e and 1f, respectively. By multi-Lorentzian fitting, the Raman peaks are in good agreement with other reports [24, 37]. Among them, the $E_{2g}^1(\Gamma)$ (~356 cm$^{-1}$) and $A_{1g}(\Gamma)$ (~ 418 cm$^{-1}$) peaks of WS$_2$ correspond to the out-of-plane vibrational mode and in-plane vibrational mode of sulfur atoms, respectively, which will be further discussed for thermal conductivity determination and mechanism analysis. One symmetric sharp PL peak is centered at ~1.97 eV, corresponding to the direct bandgap emission of monolayer WS$_2$ [8]. The above characterizations demonstrate the high crystalline quality of the CVD-synthesized monolayer WS$_2$.

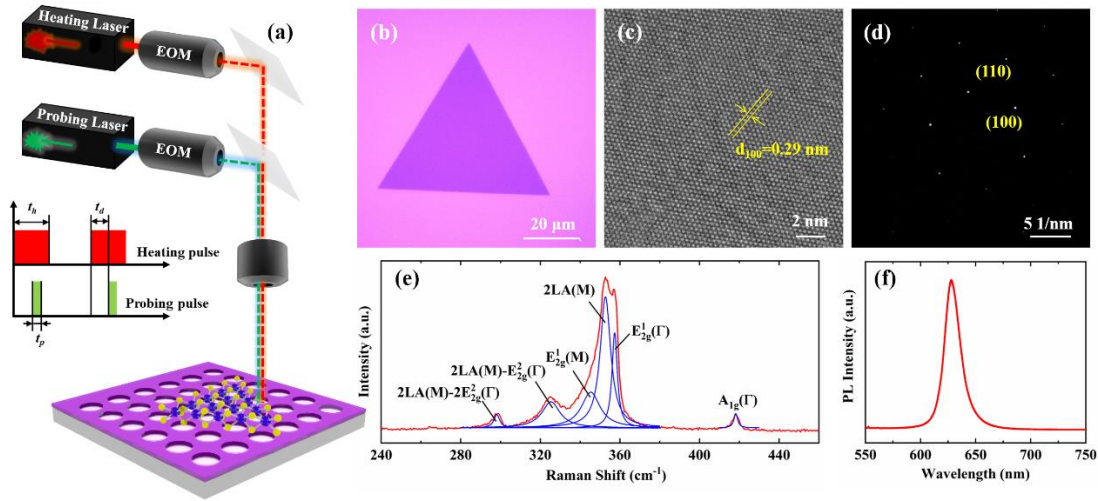

**Figure 1.** (a) Measurement system of dual-wavelength flash Raman method. (b) Optical micrograph image, (c) High-resolution transmission electron microscope (HRTEM) image, (d) Selected-area electron diffraction (SAED) pattern, (e) Raman spectrum and (f) PL spectrum of as-synthesized monolayer WS$_2$ flake on SiO$_2$/Si substrate.

To obtain the thermal conductivity of suspended and supported monolayer WS$_2$, we utilized the dual-wavelength flash Raman method developed by our previous work, which is applicable for bulk, one-dimensional (1D), 2D and 2D anisotropic materials [34, 35, 38-40]. Compared with the popular steady-state Raman spectroscopy method, this transient measurement method can eliminate the influence of laser absorption coefficient, so the measurement accuracy is significantly improved. As shown in Figure 1a, two continuous laser beams are modulated into periodic pulsed laser beams by electro-optical modulators (EOMs, ConOptics M350-160 KD*P 25D). One periodic pulsed laser beam is used to heat the sample to produce periodic temperature changes. Another periodic pulsed laser beam with different wavelength is used to detect the temperature variations by measuring the Raman peak shift. Due to the different wavelengths of the heating laser and the probing laser, the excited Raman scattering of the heating laser will not affect the acquired Raman spectrum signal based on the wavelength of probing laser. In addition, the pulse frequency of two pulsed laser beams depends on the same digital delay and pulse generator (DDPG, Stanford Research DG645) to ensure the total coincidence of the two periods. The width of heating pulse, $t_h$, and probing pulse, $t_p$, and the starting points of two pulsed laser beams (namely the pulse delay time, $t_d$) can be adjusted separately, so the complete temperature rising and dropping curves can be measured. The final results of thermal properties are determined by fitting the normalized temperature variation curves with the established 2D model.

All measurements were performed in a vacuum system, so thermal convection can be neglected. The transient heat conduction process of suspended/supported $WS_2$ within one heating and cooling period can be expressed as

$$\frac{\partial^2 \theta(r,t)}{\partial r^2} + \frac{1}{r}\frac{\partial \theta(r,t)}{\partial r} + \Phi(r) + \Omega(t) = \frac{1}{\alpha}\frac{\partial \theta(r,t)}{\partial t}$$

$$\Phi(r) = \begin{cases} \frac{\eta q}{\lambda \delta}\exp(-r^2/r_h^2) & (t \leq t_h) \\ 0 & (t \geq t_h) \end{cases} \quad (1)$$

$$\Omega(r,t) = \begin{cases} 0 & (suspended) \\ -\frac{g}{\lambda \delta}(\theta(r,t) - \theta_{sub}(r,t)) & (supportted) \end{cases}$$

where $\theta(t)$ is the temperature rise of WS$_2$ at measurement point and time $t$, $\alpha$ is the thermal diffusivity of WS$_2$, $\Phi(r)$ is the heat source term and $\Omega(t)$ is the substrate term. Among these two terms, $\eta$, $\lambda$ and $\delta$ are the effective laser absorption coefficient, the thermal conductivity and the thickness of WS$_2$, respectively. $q$ is the laser power density of the heating laser beam, $r_h$ is the laser spot radius of the heating laser beam where the power density attenuates to $q/e$, $t_h$ is the width of heating pulse, $g$ is the contact thermal conductance between WS$_2$ and the substrate, and $\theta_{sub}(t)$ is the temperature rise of the substrate surface.

To simplify the initial heat conduction equation, we utilized the nondimensional techniques and the equation 1 can be rewritten as

$$\frac{\partial^2 T(x,\tau)}{\partial x^2} + \frac{1}{x}\frac{\partial T(x,\tau)}{\partial x} + \Phi(x) + \Omega(\tau) = \frac{1}{\psi_1}\frac{\partial \theta(x,\tau)}{\partial \tau}$$

$$\Phi(x) = \begin{cases} \exp(-x^2) & (\tau \leq \tau_h) \\ 0 & (\tau \geq \tau_h) \end{cases} \quad (2)$$

$$\Omega(x,\tau) = \begin{cases} 0 & (suspended) \\ -\psi_2(1-\delta T)T(x,\tau) & (supportted) \end{cases}$$

where $x = r/r_h$ is the dimensionless coordinate with the characteristic length, $r_h$, $T = \theta/\theta_0$ is the dimensionless temperature rise of WS$_2$ with $\theta_0 = \eta q r_h^2/\lambda \delta$ as the characteristic temperature rise, and $\tau = t/t_0$, $\tau_h = t_h/t_0$ with $t_0$ as the characteristic time chosen by research to ensure the equation convergence. $\psi_1 = \alpha t_0/r_h^2$ and $\psi_2 = g r_h^2/\lambda \delta$ are two defined dimensionless numbers which influence the heat transfer process in $WS_2$ plane and between $WS_2$ and the substrate, respectively. Compared with equation 1, the temperature variation of the substrate is assumed as $\theta_{sub} = \delta T \times \theta$, where $\delta T = \theta_{sub}(0,t_h) / \theta(0,t_h)$. The uncertainty of the assumption has been verified within 1%.

By Hankel transform, the analytical solution of the temperature variation of suspended/supported $WS_2$ can be obtained as

$$T(x,\tau) = \sum_{m=1}^{\infty} T^*(\mu_m,\tau) K_0(\mu_m,x) \qquad (3)$$

where $T^*(\mu_m,\tau)$ and the kernel $K_0(\mu_m,r)$ are expressed as

$$K_0(\mu_m,r) = \frac{\sqrt{2}}{R} \frac{J_0(\mu_m r)}{J_1(\mu_m R)} \quad (R \to \infty)$$

$$T^*(\mu_m,\tau) = \begin{cases} \dfrac{E^*(\mu_m)}{\mu_m^2 + \chi_2'}\left[1 - \exp\left(-\psi_1(\mu_m^2 + \psi_2')\tau\right)\right] & (\tau \leq \tau_h) \\ T^*(\mu_m,\tau_h)\exp\left(-\psi_1(\mu_m^2 + \psi_2')(\tau - \tau_h)\right) & (\tau > \tau_h) \end{cases} \qquad (4)$$

$$E^*(\mu_m) = \int_0^R \exp(-x^2) K_0(\mu_m,x) x\, dx$$

$$\psi_2' = \psi_2(1-\delta T)\begin{cases} = 0 & (suspended) \\ \neq 0 & (supported) \end{cases}$$

where $J_0$ and $J_1$ are the zero-order and the first-order Bessel functions of the first kind, respectively. $\mu_m = \beta_m/R$ is the characteristic value, $\beta_m$ is the positive root of the first-order Bessel function and $R$ is the dimensionless characteristic length.

The measured normalized temperature rise, $\Theta$, is an average value over the pulse width and the power distribution of probing laser. When the spots of heating laser and probing laser coincide, $\Theta$ and can be expressed as

$$\Theta(\tau) = \frac{T_m(\tau)}{T_m(\tau_h)} = \frac{\int_{\tau_d}^{\tau_d+\tau_p}\int_0^{\infty} T(\tau)\exp\left(-\frac{x^2}{x_p^2}\right)xdxd\tau}{\int_{\tau_h-\tau_p}^{\tau_h}\int_0^{\infty} T(\tau)\exp\left(-\frac{x^2}{x_p^2}\right)xdxd\tau} \tag{5}$$

where $\tau_d = t_d/t_0$, $\tau_p = t_p/t_0$, $x_p = r_p/r_h$ and $x_p$ is the laser spot radius of the probing laser beam where the power density attenuates to 1/e of the center.

From the analytical solution of equation 3-5, it can be found that the normalized temperature rise, $\Theta$, is associated with the laser spot radius of the heating and probing laser beam, $r_h$ and $r_p$, and two dimensionless numbers, $\psi_1$ and $\psi_2$. Among them, $r_h$ and $r_p$ can be accurately measured, so $\psi_1$ is the only unknown for suspended case, while both $\psi_1$ and $\psi_2$ are unknown for supported case. It should be noted that $\psi_2$ can be extracted from the steady-state temperature mapping in advance. Therefore, $\psi_1$ of both suspended and supported WS$_2$ can be directly obtained by fitting the normalized temperature rise curve, $\Theta(\tau)$, and then the thermal diffusivity, $\alpha$, can be determined. With known density $\rho$ and specific heat $c_p$, the thermal conductivity, $\lambda = \alpha\rho c_p$, can be further characterized. It is worth noting that the characteristic temperature rise, $\theta_0 = \eta q r_h^2/\lambda\delta$, has been eliminated by normalization in equation 5, so the uncertainty of laser absorption coefficient of WS$_2$ will not influence the calculation results, which is a significant advantage compared with the popular steady-state Raman method.

The as-synthesized WS$_2$ were transferred onto the pre-prepared SiO$_2$/Si substrate chemically etched with an array of holes with 5 μm in diameter, as shown in Figure 2a. Before thermal properties measurement, we first calibrated the relationship between the temperature and the Raman peak shift. The temperature-dependent Raman spectra measurements of suspended and supported monolayer WS$_2$ were carried out at 100 K - 500 K under 532 nm excitation. The measured Raman spectra of supported WS$_2$ at 100 K and 500 K are shown in Figure 2b. After multiple-peak Lorentzian fitting, the obtained Raman peaks of $E_{2g}^1(\Gamma)$ and $A_{1g}(\Gamma)$ modes show a remarkable red shift with temperature increasing. In addition, the shift of both peaks versus temperature can be well fitted by a linear function as

$$\omega(T) = \chi T + \omega_0 \qquad (6)$$

where $\chi$ is the first order temperature coefficient and $\omega_0$ is the frequency of Raman peak extrapolated to 0 K. According to Figure 2c and 2d, the fitted values of slope are -0.01504 cm$^{-1}$/K ($E_{2g}^1(\Gamma)$) and -0.01433 cm$^{-1}$/K ($A_{1g}(\Gamma)$) for suspended monolayer WS$_2$, and -0.01209 cm$^{-1}$/K ($E_{2g}^1(\Gamma)$) and -0.01216 cm$^{-1}$/K ($A_{1g}(\Gamma)$) for supported monolayer WS$_2$, respectively. The results of $R^2 > 0.99$ for all fittings indicate the high goodness of fit and the measured temperature coefficients are of the same order compared with literature values, which verify the reliability of our measurement. Nevertheless, we should point out that the exact value of temperature coefficient is not important when the Raman peak shift is linearly dependent on the temperature in our method. In this case, the temperature rise is proportional to the Raman peak shift, so the scale factor can also be eliminated by normalization. Namely, the uncertainty of linear fitting, one of the main sources of errors in steady-state Raman-based method, will not affect the final measurement accuracy. Actually, the phenomenon of temperature-induced relative Raman red shift is a universally acknowledged fact arisen from the anharmonic terms in lattice potential energy from the perspective of classical thermodynamics [41] and can be fitted to the negative Bose–Einstein population from the quantum mechanical point of view [42, 43]. The phenomenon has been observed in many 2D materials such as graphene and various TMDCs. That is to say the calibration process can be omitted for common materials to simplify the measurement steps. However, the nonlinear temperature dependence of Raman shift has also been reported for several supported films mainly due to the thermal expansion effect caused by the on-site interactions with the substrates [44-46]. Based on this, the calibration procedure is necessary for supported $WS_2$ to ensure the measurement accuracy in our study.

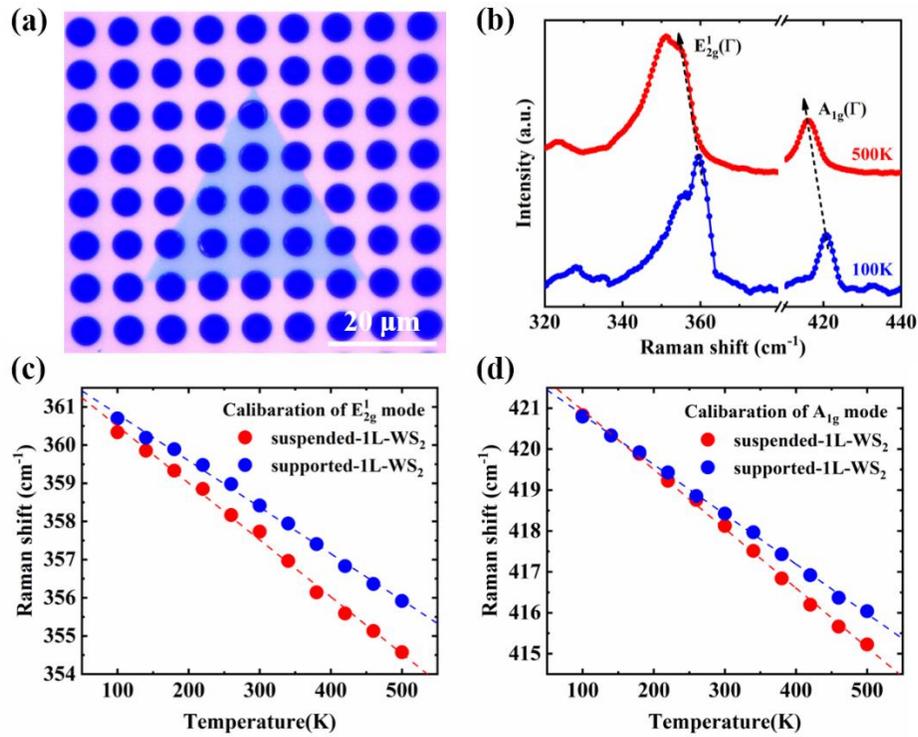

**Figure 2.** (a) Optical micrograph image of as-synthesized monolayer WS$_2$ transferred onto Si/SiO$_2$ substrate chemically etched with an array of holes with 5 μm in diameter. (b) Raman spectra of supported monolayer WS$_2$ at 100 K and 500 K for an instance. Temperature dependence of (c) E$^1_{2g}$(Γ) peaks and (d) A$_{1g}$(Γ) peak frequencies of suspended and supported monolayer WS$_2$ at 100-500 K with linear fitting.

The interfacial thermal conductance is extracted as 0.11±0.01 MW/(m$^2$·K) by steady-state temperature mapping (Figure S1 in supporting information (SI)). By fitting the normalized temperature variation curves (Figure 3b-3g) and multiplying by the specific heat and density (Table. S2, SI), the calculated thermal conductivity of suspended and supported monolayer WS$_2$ at 200 K - 400 K are shown in Figure 3a. The measured thermal conductivity of suspended monolayer WS$_2$ at ambient is 28.45±6.52 W/(m·K) which is between the result of 32 W/(m·K) reported by Peimyoo et al. [24] and 20 W/(m·K) reported by Vieira et al. [25] The slight deviation is attributed to the misestimation of laser absorption coefficient using steady-state Raman method. With temperature increasing, thermal conductivities of both suspended and supported cases show a decreasing trend as an exponential function of temperature due to the stronger influence of Umklapp phonon-phonon scattering. In addition, significant reduction in thermal conductivity can be found at wide temperature range, indicating a strong substrate effect.

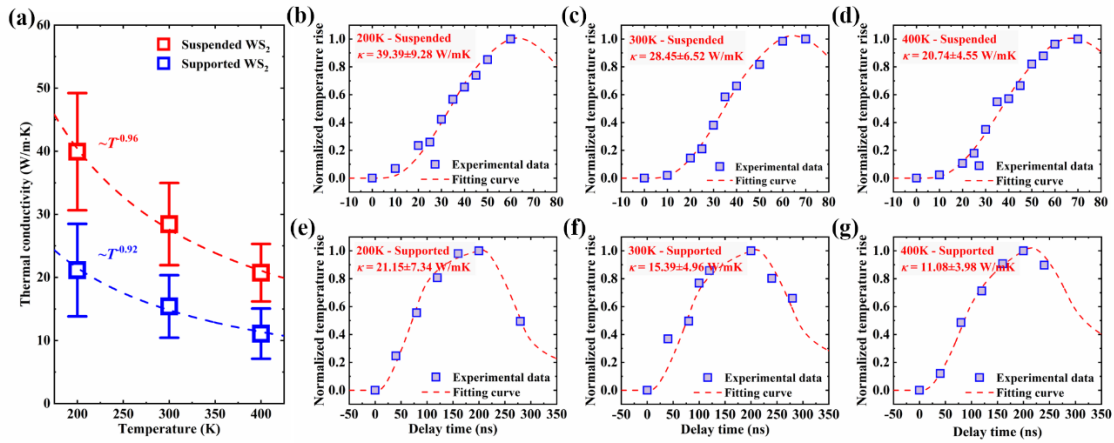

**Figure 3.** (a) Thermal conductivity of suspended and supported monolayer WS2 at 200-400 K with exponential fitting represented by dotted lines. Experimental measurement data and fitting of thermal conductivity of (b) suspended-200 K, (c) suspended-300 K, (d) suspended-400 K, (e) supported-200 K, (f) supported-300 K and (g) supported-400 K of monolayer WS2 by dual-wavelength flash Raman method.

Raman spectroscopy has been recognized to be a versatile tool to determine the materials properties, such as layer number [47], temperature variation [24, 29], defect density [48], built-in strain [49] and doping level [50, 51], and has been utilized to study the substrate effect of TMDCs supported on different substrates by Raman shift [52, 53]. Herein, we present a thorough investigation of Raman spectra by performing Raman spatial mapping from the perspective of peak shift, full with at half maximum (FWHM) and intensity ratio to study the substrate effect of monolayer $WS_2$. Considering the rich Raman spectrum of monolayer $WS_2$ excited by 532 nm, we focused exclusively on the first-order optical modes $E_{2g}^1(\Gamma)$ and $A_{1g}(\Gamma)$ at the Brillouin zone (BZ) center, as shown in Figure 4g. Figure 4a represents the schematic diagram of the Raman spatial mapping with laser of 532 nm and step of 0.5 μm, in which the mapping area contains both the suspended and supported sections. Figure 4b and 4c show the spatial images of the frequency of $\omega_{E_{2g}^1(\Gamma)}$ and $\omega_{A_{1g}(\Gamma)}$ modes, respectively, as well as the spatial maps of their FWHM shown in Figure 4d and 4e, respectively. The integrated $A_{1g}(\Gamma)$-to-$E_{2g}^1(\Gamma)$ intensity ratio $I_{A_{1g}(\Gamma)}/I_{E_{2g}^1(\Gamma)}$ is exhibited in Figure 4f. Firstly, both the $E_{2g}^1(\Gamma)$ and $A_{1g}(\Gamma)$ peaks show a blue-shift when supported on $SiO_2$/Si substrate. We attribute this stiffness to the change in doping level due to the charge transfer between $WS_2$ and substrate [51-53]. It should be noted that the strain effect is negligible after the transfer effect and the temperature effect has been eliminated by varying the incident laser power, so they will not influence the measured Raman shift. In addition, the FWHM of $E_{2g}^1(\Gamma)$ mode is broadened while that of $A_{1g}(\Gamma)$ mode becomes narrower with substrate supported. Since few studies mentioned the variation of FWHM in TMDCs, we draw an analogy between the $E_{2g}^1(\Gamma)$ mode and the G mode of graphene for they both show $E_{2g}(\Gamma)$ symmetry at BZ center [51]. Pioneering results show that the line width of the G mode is a sensitive probe of low doping levels, so the broadening FWHM of $E_{2g}^1(\Gamma)$ mode further verifies the increased doping level of supported $WS_2$. For FWHM of $A_{1g}(\Gamma)$ mode, we associate it with the $A_{1g}(\Gamma)$ phonon lifetime based on the energy-time uncertainty relationship [54] as

$$\frac{\Gamma}{\hbar} = \frac{1}{\tau} \qquad (7)$$

where $\tau$ is the phonon lifetime (ps), $\hbar$ is the modified Planck constant ($5.3\times10^{-12}$ cm$^{-1}$ s), and $\Gamma$ is the measured FWHM (cm$^{-1}$). The obtained averaged $A_{1g}(\Gamma)$ phonon lifetime of suspended and supported monolayer $WS_2$ are 1.32 ps and 0.66 ps, respectively, which is consistent with the calculation results from first-principles study in the order of magnitude [55]. It is clear that $A_{1g}(\Gamma)$ phonon lifetime is reduced due to phonon-substrate scattering, thus resulting in the reduction in thermal conductivity. At last, the intensity ratio between $A_{1g}(\Gamma)$ mode and $E_{2g}^1(\Gamma)$ mode ($I_{A_{1g}(\Gamma)}/I_{E_{2g}^1(\Gamma)}$) is significantly weakened in supported $WS_2$. As $A_{1g}(\Gamma)$ mode and $E_{2g}^1(\Gamma)$ mode are representative to the out-of-plane and in-plane vibrational modes, respectively, the Raman intensity of $E_{2g}^1(\Gamma)$ mode remains constant while that of $A_{1g}(\Gamma)$ mode is largely suppressed due to substrate effect. Therefore, from the Raman spectra analysis we can conclude that the increase of doping level, reduction of phonon lifetime and suppression of out-of-plane vibration mode caused by the substrate effect lead to the decrease in thermal conductivity.

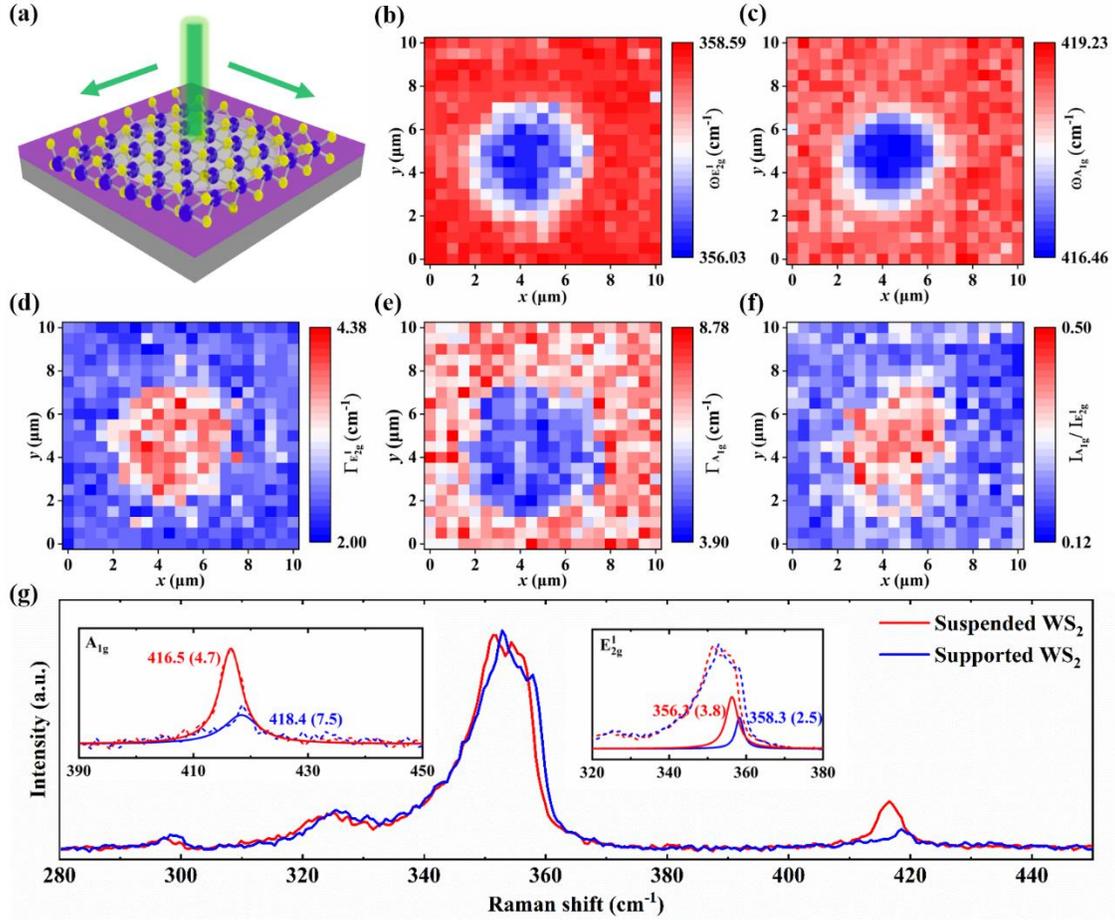

**Figure 4.** (a) Schematic diagram of the Raman spatial mapping with laser of 532 nm and step of 0.5 μm. Raman spatial mapping images of (b) $\omega_{E_{2g}^1(\Gamma)}$, (c) $\omega_{A_{1g}(\Gamma)}$, (d) $\Gamma_{E_{2g}^1(\Gamma)}$, (e) $\Gamma_{A_{1g}(\Gamma)}$ and (f) $I_{A_{1g}(\Gamma)}/I_{E_{2g}^1(\Gamma)}$. (g) Normalized Raman spectra recorded for suspended (red solid line) and supported (blue solid line) areas in sample (a), respectively. The local enlarged view of $E_{2g}^1(\Gamma)$ and $A_{1g}(\Gamma)$ mode with frequency and FWHM labeled are shown for detailed comparison.

In parallel to the experimental measurements, we conduct the computational simulation to deepen our understanding. Full-atom nonequilibrium molecular dynamics (NEMD) simulation was implemented to model the suspended and supported monolayer WS$_2$, as depicted in Figure 5a. The system sizes are set as 50.5 nm in length and 5.9 nm in width based on the fact that the thermal conductivity convergences at 5 nm with the increasing of width [56, 57]. Thermal baths are applied in length direction to calculate the thermal conductivity at different temperatures. As demonstrated in Figure 5b, both decrease trends are observed in thermal conductivity with temperature increase and substrate supported, which are in consistent with measurement results. The ~$T^{-\beta}$ relation of thermal conductivity for both suspended and supported cases reflect the stronger Umklapp phonon-phonon scattering with temperature increasing, which is consistent with the measurement results (Figure 3a). In addition, suspended WS$_2$ has a $T^{-0.68}$ behavior, which is close to a Slack relation ($\beta = 1$) [58]. The slight deviation is deduced by the boundary scattering since the longitudinal length studied here is much smaller than phonon mean free path (MFP). However, supported WS$_2$ shows a smaller power exponent with relationship of $T^{-0.57}$. Such reduction of $\beta$ has been observed in supported CNT and is attributed to the suppression of long-range acoustic phonons due to the substrate interaction [56].

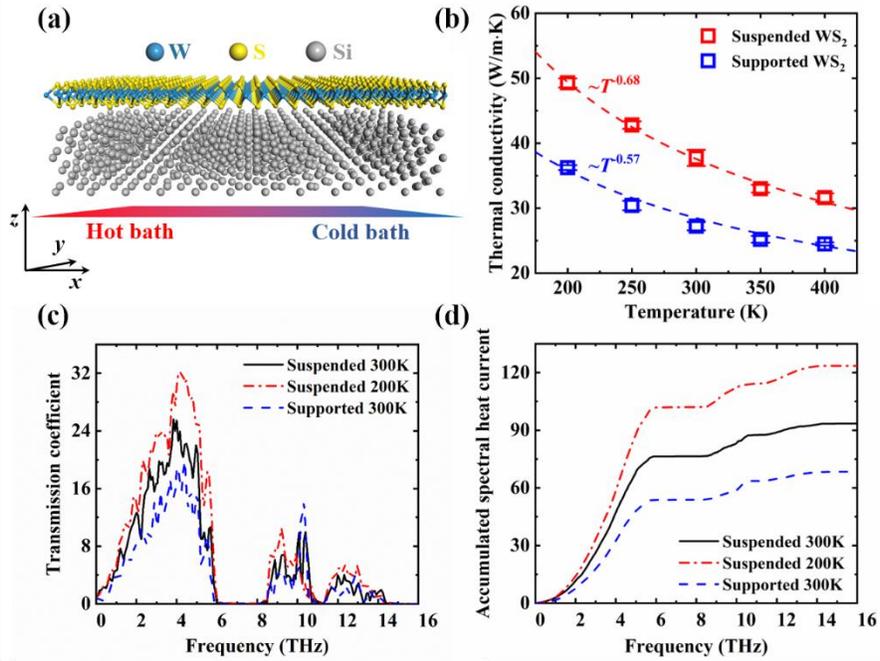

**Figure 5.** (a) The schematic diagram of NEMD simulation configuration for calculating the thermal conductivity of supported monolayer WS$_2$. (b) Thermal conductivity of suspended and supported monolayer WS$_2$ versus temperature at 200-400 K. The dotted lines draw the exponential dependence $\kappa \sim T^{-\beta}$. (c) Spectral transmission coefficient and (d) accumulated spectral heat current for monolayer WS$_2$ of suspended-300 K, suspended-300 K and supported-300 K, respectively.

To further investigate the underlying mechanism of the substrate effect, we perform the calculations of spectral phonon transmission coefficient, including full orders of phonon-phonon interactions using the method developed by Sääskilahti et al [59, 60]. In Figure 5c, it can be found that phonons share the same range of frequency from 0 THz to 15 THz with a significant band gap around 6-8 THz for three selected cases. With temperature increasing or substrate supported, the transmission coefficient of acoustic phonons is reduced more pronouncedly than that of optical phonons, similar to the widespread study of two-dimensional graphene sheet [61, 62]. This phenomenon can be reflected more clearly by the contribution of phonons of different frequencies to thermal conductivity (in Figure 5d), which is obtained by integrating the transmission coefficient. For temperature effect, the transmission coefficient of suspended-300 K is significantly lower than that of suspended-200 K mainly in range from 2.5 THz to 5 THz due to enhanced phonon scattering. However, the transmission coefficient is reduced uniformly in all range of acoustic phonons for substrate effect. In addition, the reduction of transmission coefficient around 12 THz (corresponds to the ZO branch) for supported $WS_2$ further validates the reduction of phonon lifetime and suppression of $A_{1g}(\Gamma)$ mode in Raman spectra analysis. Therefore, the phonon-substrate scattering induced by the $WS_2$-substrate interaction suppresses the acoustic phonons, which mainly accounts for the attenuation of the thermal conductivity of supported monolayer $WS_2$.

Based on the Raman spectra analysis and computational simulation, we have confirmed the decreasing trend with substrate supported and revealed the underlying mechanism from both acoustic and optical phonons. To extent our study, we reviewed the experimental measurements to date on the thermal conductivity of the suspended and supported monolayer TMDCs at room temperature to show the substrate effect (in Table 1). It can be found that the substrate effect significantly influence the thermal conductivity of the listed 2D TMDCs with reduction of 25%~60%. For the discussed monolayer TMDCs ($MX_2$, M = Mo, W, X = S, Se), they share the same planar structure with one layer of M (transition metal) atoms sandwiched between two layers of X (chalcogen) atoms. The same crystal structure and similar atomic mass leads to the similar phonon dispersion and density of states [63], which means the similar modes of phonon transports. Therefore, we think our research of monolayer $WS_2$ can be further extended to help understand the substrate effect on thermal properties of other TMDCs.

**Table 1.** Comparison of the existing experimental data of the conductivity of suspended and supported monolayer TMDCs at 300 K. The measurement methods are all (refined) stead-state Raman method except for the transient Raman method in our work.

| Material | Thermal conductivity (W/mK) | | Interfacial thermal conductance (MW/m²K) | Reference |
|---|---|---|---|---|
| | suspended | supported | | |
| $MoS_2$ | 34.5±4 | | | 64 |
| | 13.3±4 | | | 65 |
| | | 62.2[a] | 1.94 | 28 |
| | 84±17 | 55±20[b] | 0.44±0.07 | 29 |
| $WS_2$ | 32 | | | 24 |
| | 20 (16) | | | 25 |
| | 28.45±6.52 | 15.39±4.96[a] | 0.11±0.01 | This work |
| $MoSe_2$ | 59±18 | 24±11[a] | 0.09±0.03 | 29 |
| $WSe_2$ | 49±14 | 37±12[a] | 2.95±0.46 | 29 |

[a] sample on $SiO_2/Si$, [b] sample on Au

CONCLUSION

In this paper, the dual-wavelength flash Raman method is used to experimentally measure the thermal conductivity of the suspended and the Si/SiO$_2$ substrate supported monolayer WS$_2$ at a temperature range of 200 - 400 K. The measurement results show that as the temperature increases, the thermal conductivity show a downward trend, and the thermal conductivity of supported WS$_2$ is much lower than that of the suspended one at different temperatures. The Raman spatial mapping shows that the substrate increase the doping concentration, leading to a hardening effect and reduced phonon lifetime of WS$_2$. The thermal conductivities of the suspended and supported monolayer WS$_2$ under the same conditions were calculated by MD simulation, and the calculated results are consistent with the experimental trend. In addition, the phonon transmission coefficient was mutually verified with Raman spectra analysis and further revealed the underlying mechanism that the substrate effect significantly enhances the phonon scattering at the interface and mainly suppresses the acoustic phonon, thus leading to the reduction of thermal conductivity. We believe our work is of great significance to the understanding of the substrate effect of 2D TMDCs and can provide guidance for thermal management of TMDCs-based devices.

EXPERIMENTAL AND COMPUTATIONAL SECTION

**Sample growth and transfer process.** The monolayer WS$_2$ sample was grown by AP-CVD route. Sodium tungstate (Na$_2$WO$_4$) powder and selenium powder were used as W and Se precursors, respectively. Na$_2$WO$_4$ powder was dissolved in water, forming 1.8 mg/mL solution. For the synthesis of monolayer WS$_2$, 5 μL solution was spin-coated onto the SiO$_2$/Si substrate at a speed of 3000 rpm for 60 s. Then, the precursor-coated substrate was placed in a quartz boat and loaded in the center of the furnace, another quartz boat containing ~ 950 mg Se powder was placed at the upstream of the furnace. Before growth, the quartz reactor was purged with large Ar gas flow for ~ 5 min. Subsequently, the furnace was heated up to 800 °C at a speed of 20 °C/min with a 80 sccm Ar gas, then kept at this temperature for 10 min with the mixture gas of Ar/H2 (80/6 sccm/sccm) to grow monolayer WS$_2$. After the growth, the quartz reactor was pulled out of the furnace to rapidly cool down to the room temperature.

For the transfer of WS$_2$ crystals, the WS$_2$/SiO$_2$/Si was first spin-coated with PMMA at a speed of 2000 rpm for 60 s, followed by curing at 180 °C for 90 s. Then, the PMMA/ WS$_2$/SiO$_2$ was immersed into 2M KOH solution to etch the SiO$_2$ layer. Subsequently, the PMMA/WS$_2$ layer was delaminated from SiO$_2$/Si substrate, followed by fishing onto the porous SiO$_2$/Si substrate with the diameter of 5 μm. Finally, the PMMA layer was removed by immersing the above substrate into acetone and dried by critical point dryer (Leica EM CPD300) with supercritical carbon dioxide to avoid the damage of monolayer WS$_2$.

**Characterizations of as-synthesized monolayer WS$_2$.** The micro-Raman and PL characterization were carried out by HORIBA LabRAM HR Evolution Raman system with 532 nm excitation laser and a 50× objective lens with numerical aperture (NA) of 0.55. To avoid heating induced by the excitation laser, the laser power was maintained at ~0.1 mW. Optical images were performed by an Olympus BX53 microscope. TEM and SAED images were carried out in a JEM-2100 system at an acceleration voltage of 200 kV

**Calibration and thermal conductivity measurement.** In temperature-dependent and thermal conductivity measurement, the sample was set inside the vacuum chamber (Linkam, THMS350EV-4) to control the temperature with temperature controlling accuracy of 0.01 K and maintain a vacuum of ~$10^{-4}$ Pa by a two-stage hybrid pump set (a scroll vacuum pump (Leybold SCROLLVAC 7 plus) and a turbomolecular vacuum pump (Leybold TURBOVAC 90 i)). A 40× objective lens with NA of 0.6 and an adjustable focal length is used to achieve good focus after the optical window. The heating laser beam was excited by laser with 633 nm to avoid damage to the sample and the probing laser beam was excited by laser with 532 nm. By moving the laser across a sharp edge and fitting the Raman profile by a Gaussian function, the laser spot radius of heating laser and probing laser beams are $r_h$ = 0.34 μm and $r_p$ = 0.29 μm, respectively, which is close to the numerical aperture estimation [27]. In transient measurement, the width of heating pulse for suspended $WS_2$ is $t_h$ = 50 ns (with $t_p$ = 30 ns) to satisfy the semi-infinite hypothesis, in which the heat has not been transferred to the substrate and the influence of thermal contact resistance can be eliminated (Figure S2, SI). The width of heating pulse and probing laser for supported $WS_2$ is $t_h$ = 200 ns and $t_p$ = 80 ns, respectively. The laser power of heating laser was adjusted to control the maximum transient temperature rise of ~ 50 K according to the thermal properties of the $WS_2$ sample at different temperature. The laser power of probing laser was ~0.6 mW to excite the strong enough signal and not to damage the sample.

**NEMD simulation process.** In the simulation, the Stillinger-Weber (SW) potential parameterized by Jiang et al. [66] is employed to describe the covalent interaction between W-S atoms. This SW potential can successfully reproduce the phonon dispersions that agree well with the first-principle results and is widely used in MD simulation of TMDCs [67, 68]. Periodic boundary condition is applied in width direction to eliminate the edge effect and atoms at two ends of length direction is fixed in space. A 10 nm vacuum layer is added in $z$ direction to avoid the atomic interaction with the system image. Langevin heat baths [69] are applied at atoms adjacent to the fixed ends to form a temperature gradient with high temperature maintained at $T_h = T_0+30$ K and low temperature at $T_c = T_0-30$ K, where $T_0$ is the environment temperature. The velocity-Verlet algorithm is used to integrate the differential equations of motions with a time step of 0.5 fs for all cases. Before NEMD simulations, the initial system is relaxed in the isothermal-isobaric (NPT) ensemble for 0.5 ns to release the internal stress. Then the system is placed under Nosé-Hoover thermostat to reach equilibrium at the designated temperature of $T_0$. After relaxing in the canonical ensemble (NVT) for 0.5 ns, the system is then switched to the microcanonical ensemble (NVE) for another 0.5 ns. Following equilibration, the thermal baths are conducted to perform the NEMD simulation for long enough time of 6 ns to ensure that the steady state is achieved. In-plane thermal conductivity is calculated according to the Fourier law, as

$$k = -\frac{J/A}{dT/dx} \tag{8}$$

where A is the cross-sectional area, defined as the product of width and thickness. The thickness of monolayer $WS_2$ is chosen as 6.12 Å which has been used in previous calculations [70]. $J$ and $dT/dx$ are the heat flux transported in the system and temperature gradient, respectively. The heat flux is recorded by the average of the input/output power at the two baths and calculated as

$$J = \frac{\Delta E_{in} + \Delta E_{out}}{2\Delta t} \qquad (9)$$

where $\Delta E_{in}$ and $\Delta E_{out}$ are the energy added and removed from each bath at each time step $\Delta t$. Both the heat flux and temperature gradient are extracted by the average value over the last 1 ns simulations when the temperature gradient is well established and the heat current is independent of simulation time, as shown in Figure S3a and S3b. All the results in this work are ensemble averaged over six independent runs with different initial conditions.

ASSOCIATED CONTENT

**Supporting Information:** Experimental uncertainty analysis; Fitting of the transient and steady-state temperature variation curves; Feasibility analysis of semi-infinite model; Well-established temperature gradient and heat current.

AUTHOR INFORMATION

**Corresponding Author**

Xing Zhang, E-mail: x-zhang@mail.tsinghua.edu.cn

Ruitao Lv, E-mail: lvruitao@tsinghua.edu.cn

**Author Contributions**

The manuscript was written through contributions of all authors. All authors have given approval to the final version of the manuscript. † Y. Z. and Q. L. contributed equally to this work.

**Notes**

The authors declare no competing financial interest

ACKNOWLEDGEMENTS

This work was supported by the National Natural Science Foundation of China (Grant Nos. 51636002, 51827807, 51972191)